\documentclass[twocolumn,superscriptaddress,amsmath,amssymb,aps,pra]{revtex4}
\usepackage{amsmath,amsthm,amssymb}
\usepackage{latexsym}
\usepackage{amscd,graphicx}
\usepackage[titletoc]{appendix}
\usepackage{epsfig}
\usepackage{epstopdf}
\usepackage[normalem]{ulem}
\usepackage[dvipsnames]{xcolor}
\usepackage[colorlinks=true,linkcolor=red,citecolor=red]{hyperref}

\newcommand{\abs}[1]{| #1 |}

\begin{document}
	\title{Higher-order exceptional point in a pseudo-Hermitian cavity optomechanical system}
	
	\author{Wei Xiong}
	\affiliation{Department of Physics, Wenzhou University, Zhejiang 325035, China}
	\affiliation{Interdisciplinary Center of Quantum Information and Zhejiang Province Key Laboratory of Quantum Technology and Device, Department of Physics and State Key Laboratory of Modern Optical Instrumentation, Zhejiang University, Hangzhou 310027, China}
	
	\author{Zhuanxia Li}
	\affiliation{Department of Physics, Wenzhou University, Zhejiang 325035, China}
	
	\author{Yiling Song}
	\affiliation{Department of Physics, Wenzhou University, Zhejiang 325035, China}
	
	\author{Jiaojiao Chen}
	\altaffiliation{jjchenphys@hotmail.com}
	\affiliation{Hefei Preschool Education College, Hefei 230013, China}
	
	\author{Guo-Qiang Zhang}
	\altaffiliation{zhangguoqiang@csrc.ac.cn}
	\affiliation{Interdisciplinary Center of Quantum Information and Zhejiang Province Key Laboratory of Quantum Technology and Device, Department of Physics and State Key Laboratory of Modern Optical Instrumentation, Zhejiang University, Hangzhou 310027, China}
	
    \author{Mingfeng Wang}
    \altaffiliation{mfwang@wzu.edu.cn}
	\affiliation{Department of Physics, Wenzhou University, Zhejiang 325035, China}
	
	
	\date{\today }
	
	\begin{abstract}
		Higher-order exceptional points (EPs), resulting from non-Hermitian degeneracies, have shown greater advantages in sensitive enhancement than second-order EPs (EP2s). Therefore, seeking higher-order EPs in various quantum systems is important for quantum information science. Here we propose a benchmark cavity optomechanical (COM) system consisting of a mechanical resonator coupled to two cavities via radiation pressure for predicting the third-order exceptional point (EP3). We first give the pseudo-Hermitian condition for the non-Hermitian COM system by taking the bath effects into account. Then we consider the mechanical gain effect and we find that the pseudo-Hermitian COM system without $\mathcal{PT}$ symmetry can host both  EP3 and EP2 for symmetric and asymmetric cavities. In the symmetric case, only EP3 or EP2 can be predicted in the parameter space, but EP3 and EP2 can be transformed into each other by tuning the optomechanical coupling strength in the asymmetric case. We further consider the case of one cavity with gain. For this case, the pseudo-Hermitian COM system is $\mathcal{PT}$ symmetric and can also host EP3 or EP2. The influence of system parameters on them is discussed. Our proposal provides a potential way to realize sensitive detection and study other physical phenomena {around} higher-order EPs in non-Hermitian COM systems.
	\end{abstract}
	
	\maketitle
	
\section{Introduction}

  Recently, cavity optomechanical (COM) systems  have emerged as a field with numerous exciting prospects for fundamental science and applications~\cite{Aspelmeyer}. Generically, typical optomechanical systems are formed by a cavity with two mirrors (one is
  fixed and the other is movable) and a mechanical resonator (MR). The radiation pressure proportional to the cavity photon number acting on the MR causes the movable mirror to vibrate. This vibration in turn changes the length of the cavity (and thus the frequency of the cavity modes) and gives rise to
  a nonlinearly optomechanical coupling between the cavity and the mechanical modes. Such an unconventional interaction results in rich backaction effects including sensing~\cite{Schreppler-2014,Wu-2017,Gil-Santos-2020,Fischer-2019}, ground-state cooling~\cite{Chan-2011,Teufel-2011}, squeezed light generation~\cite{Purdy-2014,Safavi-Naeini-2013,Aggarwal-2020}, nonreciprocal transport~\cite{Xu-2019,Shen-2016}, optomechanically induced transparency~\cite{Kronwald-2013,Weis-2010,Liuy-2013}, coupling enhancement~\cite{Xiong-2021,Lu-2013,Xiong2-2021}, and nonlinear behaviors ({e.g., bi- and tristability and chaos} )~\cite{Lu-2015,Xiong3-2016}.

  In addition, the realistic COM system is always inevitably coupled to its surrounding environment, leading to the system decoherence~\cite{Weiss-2012}.  When considering its decoherence, the open quantum system can be effectively described by a non-Hermitian Hamiltonian $H$, which violates the Hermitian condition (i.e., $H\neq H^\dag$)~\cite{Minganti-2019,Zhang-2021,Ozdemir-2019}. Generally speaking, the non-Hermitian Hamiltonian $H$ does not have real eigen energies. However, if a non-Hermitian Hamiltonian $H$ satisfies the pseudo-Hermitian condition $UHU^{-1}=H^\dag$ with $U$ being a linear Hermitian operator, its eigen energies can be either real or complex-conjugate pairs~\cite{Mostafazadeh1-2002,Mostafazadeh2-2002}. The parity-time ($\mathcal{PT}$)-symmetry Hamiltonian, $[H,\mathcal{PT}]=0$, is a special subset of pseudo-Hermitian Hamiltonians~\cite{Konotop-2016}. By tuning one parameter of a pseudo-Hermitian system, the quantum phase transition from the $\mathcal{PT}$ symmetric phase with real eigen energies to the $\mathcal{PT}$ symmetric broken phase with complex-conjugate pairs eigen energies can occur at the second-order exceptional point (EP2), where the two eigenvalues and the corresponding eigenvectors simultaneously coalesce~\cite{Bender-2013,LYL-2017,Parto-2021}.  During the past few years, EP2s have been widely investigated in COM systems~\cite{Jing-2014,Xu-2016,LYL-2017,ZhangJQ-2021,XuH-2021,Xuxw-2015} and other systems including{, e.g.,} waveguides~\cite{Doppler-2016}, microcavities~\cite{Chang-2014}, cavity magnonics~\cite{Zhang-2017,Harder-2017} and superconducting circuits~\cite{Quijandria-2018,Zhang-2019-2,Naghiloo-2019}. Around EP2s, lots of fascinating phenomena like unidirectional invisibility~\cite{Peng-2014,Lin-2011,Chang-2014}, single-mode lasing~\cite{Feng-2014,Hodaei-2014}, sensitivity enhancement~\cite{Chen-2017,Hokmabadi-2019}, energy harvesting~\cite{Fern-2021}, protecting the classification of exceptional nodal topologies~\cite{Marcus-2021}, and electromagnetically induced transparency~\cite{Guo-2009,Wang-2019,Wang1-2020,Lu-2021} can be realized.
  Besides EP2s, non-Hermitian systems can also host higher-order EPs, where more than two
  eigenmodes coalesce into one~\cite{Graefe-2008,Heiss-2008,Demange-2012,Heiss-2016,Jing-2017,Ge-2015,Ding-2016,Lin-2016,Quiroz-2019}. Very recently, the fourth-order EP was demonstrated using optical elements~\cite{Bian-2020}. It has been shown that higher-order EPs can exhibit greater advantages than EP2s in sensitive detection~\cite{Hodaei-2017,Zeng-2021,Wang-2021,Zeng-2019}, topological characteristics~\cite{Ding-2016,Delplace-2021,Mandal-2021} and spontaneous emission enhancement~\cite{Lin-2016}. With these superiorities, higher-order EPs are being intensively studied in various systems~\cite{Roy-2021,Zhong-2020,Zhang-2020,Pan-2019,Zhang-2019,Kullig-2019,Kullig-2018,Schnabel-2017,Nada-2017,Wang-2020} but attract less attention in pseudo-Hermitian COM systems. For this, seeking higher-order EPs in pseudo-Hermitian COM systems is strongly demanded since they may provide a new perspective to study conventional phenomena in COM systems ~\cite{Schreppler-2014,Wu-2017,Gil-Santos-2020,Fischer-2019,Chan-2011,
  Teufel-2011,Purdy-2014,Safavi-Naeini-2013,Aggarwal-2020,Xu-2019,Shen-2016,Kronwald-2013,Weis-2010,Liuy-2013,
  Xiong-2021,Lu-2013,Xiong2-2021,Xiong3-2016,Lu-2015}.

  In this paper, we theoretically propose a pseudo-Hermitian COM system consisting of two cavities coupled to a common MR via radiation pressure to predict the third-order exceptional point (EP3).
  First, we derive an effective non-Hermitian Hamiltonian for the proposed COM system and analytically give the pseudo-Hermitian condition of the non-Hermitian Hamiltonian in the general case. Then, two scenarios are specifically considered in the pseudo-Hermitian condition: (i) the MR is active, but the two cavities are passive; (ii) one of cavities is active, but the other cavity and the MR are both passive.
  In case (i), the proposed pseudo-Hermitian COM system without $\mathcal{PT}$ symmetry can host EP3 or EP2. We show only EP3 and EP2 can be predicted in parameter space when two symmetric cavities (i.e., two cavities have the same loss rates) are considered. But when two cavities are asymmetric (i.e., two cavities have different loss rates), not only can EP3 or EP2 be predicted, but also the transformation between EP3 and EP2 can be achieved by tuning the system parameters such as the optomechanical coupling strength. But for the case (ii), the pseudo-Hermitian condition can be further reduced to the $\mathcal{PT}$ symmetric condition by neglecting the negligible mechanical loss compared to the cavity loss rate. We show such a $\mathcal{PT}$ symmetric system can host EP3 as well as the EP2 in parameter space. We also investigate the effects of the system parameters such as the frequency detuning and optomechanical coupling strength on EP3 and EP2. Our proposal provides a promising path to engineer the pseudo-Hermitian COM system with or without $\mathcal{PT}$ symmetry for predicting higher-order EPs.

  This paper is organized as follows. In Sec. II, the model is described and the system effective Hamiltonian is given.  Then we derive the pesudo-Hermitian condition for the considered non-Hermitian COM system in Sec. III. In Sec. IV, EP3 and EP2 are studied using the pseudo-Hermitian COM system without $\mathcal{PT}$ symmetry. In Sec. V, EP3 and EP2 are studied using the pseudo-Hermitian COM system with $\mathcal{PT}$ symmetry, and the impacts of the system parameters on the EP3 and EP2 are discussed. Finally, a conclusion is given in Sec. VI. A list of
  symbols and abbreviations is given in Table I.

\begin{figure}
	\center
	\includegraphics[scale=0.36]{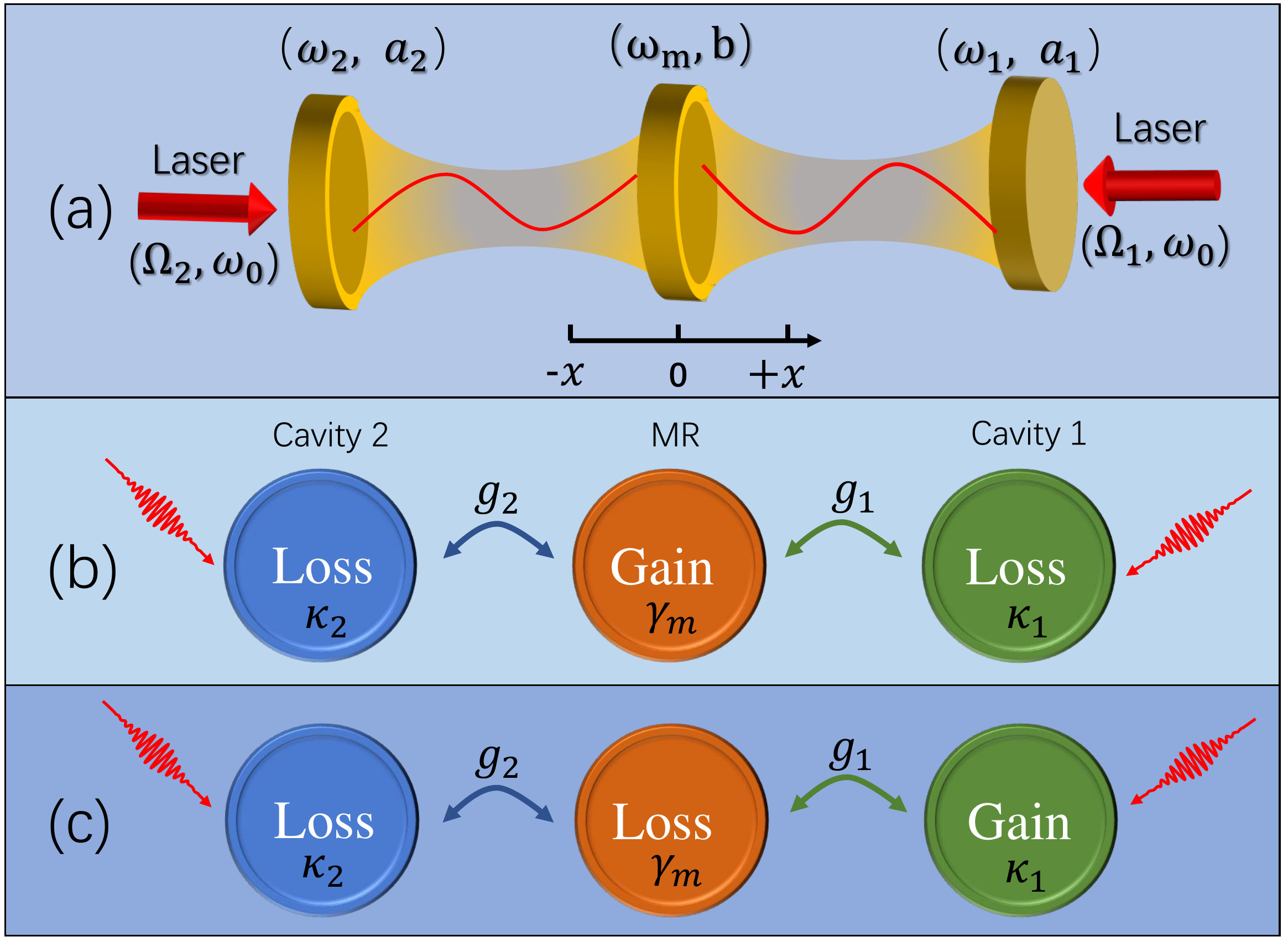}
	\caption{(a) Schematic diagram of the proposed COM system consisting of two cavities with frequencies $\omega_1$ and $\omega_2$ coupled to a common MR with frequency $\omega_m$. The two cavities are driven by two laser fields with the same frequency $\omega_0$. The corresponding amplitudes are $\Omega_1$ and $\Omega_2$. (b) The MR with gain and both the cavities with loss are considered. (c) Cavity 1 with gain and both the MR and cavity 2 with loss are considered. In both (b) and (c), $\kappa_1,~\kappa_2$ and $\gamma_m$ are gain or loss rates for two cavities and the MR. $g_{1(2)}$ is the single-photon optomechanical coupling strength between cavity 1~(2) and the MR.}
	\label{fig1}
\end{figure}
\begin{figure}
	\center
	\includegraphics[scale=0.6]{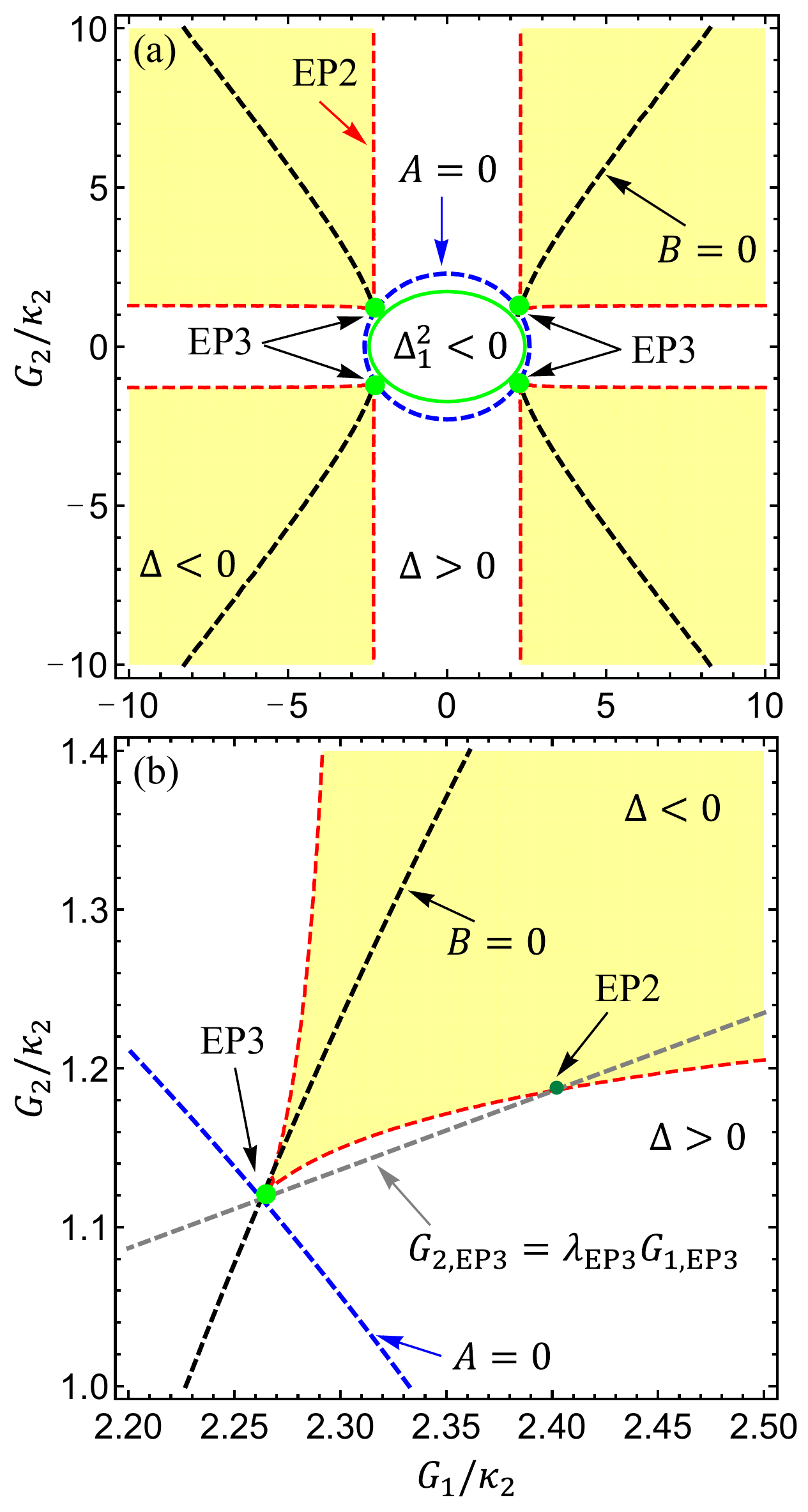}
	\caption{(a) The phase diagram of the discriminant given by Eq.~(\ref{s11}) with $\gamma_m=-(\kappa_1+\kappa_2)<0$ vs the normalized parameters $G_1/\kappa_2$ and $G_2/\kappa_2$. (b) The ranges of $2.2\leq G_1/\kappa_2\leq 2.5$ and $1.0\leq G_2/\kappa_2\leq 1.4$ are plotted to show the EP2 along the red curve. One EP2 is predicted at the crossing point of the red and gray curves.}
	\label{fig2}
\end{figure}
	
\section{Model}

We consider a COM system consisting of two cavities (labeled as cavity $1$ and cavity $2$) with angular frequencies $\omega_1$ and $\omega_2$ radially coupled to a common MR with frequency $\omega_m$~(see Fig.~\ref{fig1}), where two cavities are subjected to two strong laser fields with the same frequency $\omega_0$. This setup was achieved in several experiments~\cite{Dong-2012,Hill-2012,Andrews-2014}. In the rotating frame of the laser field, the total Hamiltonian of the COM system reads (setting $\hbar=1$)~\cite{Zhangk-2015}
\begin{align}
H=&\delta_1 a_1^\dag a_1+\delta_2 a_2^\dag a_2+\omega_m b^\dag b\nonumber\\
&-(g_1 a_1^\dag a_1-g_2 a_2^\dag a_2) (b^\dag+b)\nonumber\\
&+i\Omega_1(a_1^\dag-a_1)+i\Omega_2(a_2^\dag-a_2),\label{q1}
\end{align}
where $\delta_{1(2)}=\omega_{1(2)}-\omega_0$ is the frequency detuning of cavity $1$ ($2$) from the laser field. $g_1$ and $g_2$ are the single-photon optomechanical coupling strengths. $a_{1(2)}$ and $a_{1(2)}^\dag$ are the annihilation and creation operators of cavity $1$ ($2$). $b$ and $b^\dag$ are the annihilation and creation operators of the MR. The normalized amplitudes of the laser fields to the photon flux at the inputs of cavities $1$ and $2$ are respectively $\Omega_1=\sqrt{P_1\kappa_1/\hbar\omega_1}$ and $\Omega_2=\sqrt{P_2\kappa_2/\hbar\omega_2}$, where $P_{1}$ and $P_{2}$ correspond to the powers of two laser fields, and $\kappa_1$ and $\kappa_2$ denote the decay rates of cavities $1$ and $2$.
\begin{figure}
	\center
	\includegraphics[scale=0.35]{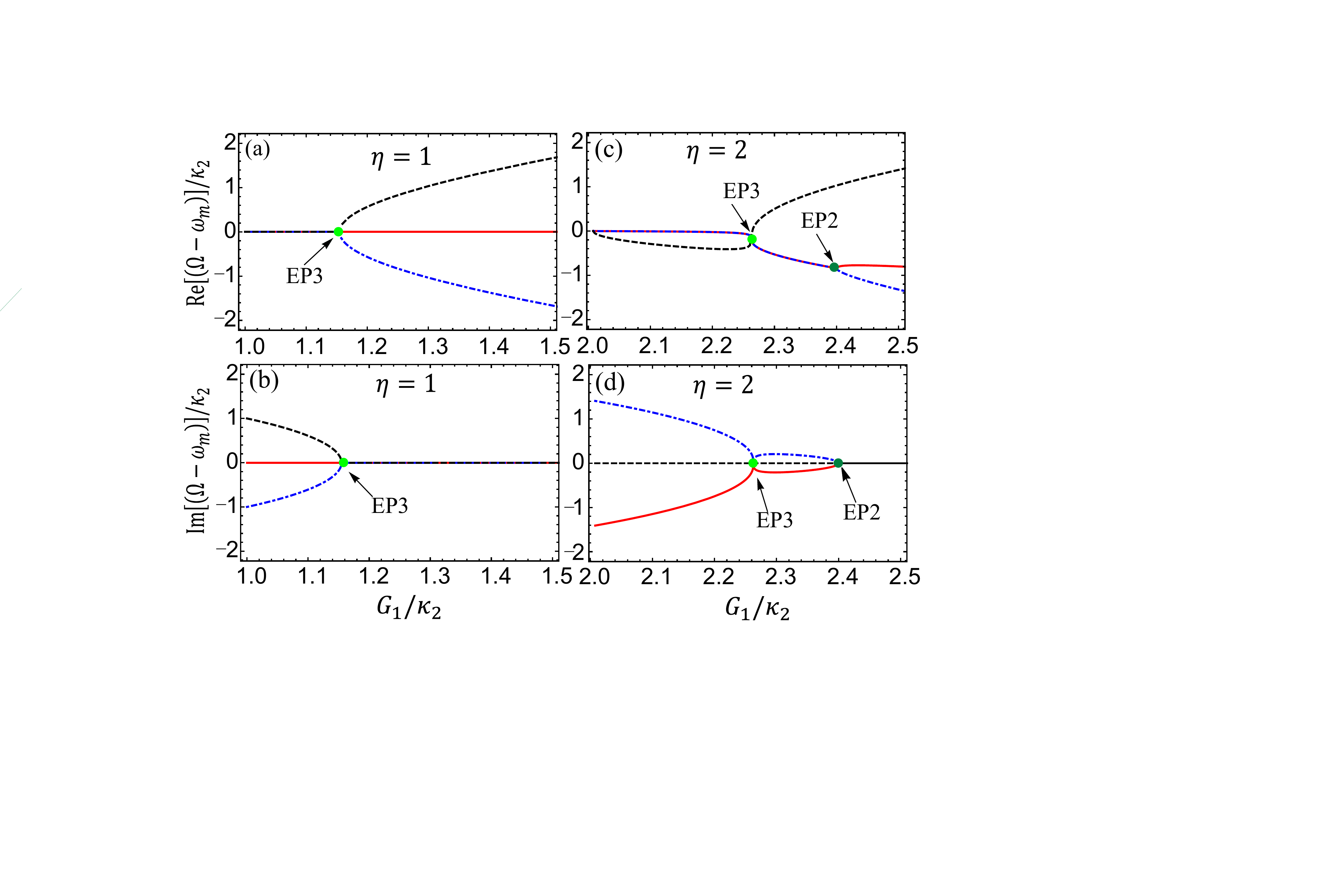}
	\caption{The real and imaginary parts of the eigenvalues given by Eq.~(\ref{q12}) with $\gamma_m=-(1+\eta)\kappa_2<0$ vs the normalized parameter $G_1/\kappa_2$. In (a) and (b), $\eta=1$ corresponds to the symmetric case. In (c) and (d), $\eta=2$ corresponds to the asymmetric case.}
	\label{fig3}
\end{figure}
Using the quantum Langevin equation approach~\cite{Benguria-1981}, the dynamics of the composite system can be expressed as
\begin{align}\label{s1}
\dot{a}_1=&-(i\delta_1+\kappa_1)a_1+ig_1 a_1 (b^\dag+b)+\Omega_1+\sqrt{2\kappa_1}a_{\rm 1,in},\nonumber\\
\dot{a}_2=&-(i\delta_2+\kappa_2)a_1-ig_2 a_2 (b^\dag+b)+\Omega_2+\sqrt{2\kappa_2}a_{\rm 2,in},\nonumber\\
\dot{b}=&-(i\omega_m+\gamma_m)b+i(g_1 a_1^\dag a_1-g_2 a_2^\dag a_2)+\sqrt{2\gamma_m}b_{\rm in},
\end{align}
where $\gamma_m$ is the decay rate of the MR. $a_{\rm 1(2),in}$ and $b_{\rm in}$ are the input noises with zero expectation value, i.e., $\langle a_{\rm 1,in}\rangle=\langle a_{\rm 2,in}\rangle=\langle b_{\rm in}\rangle=0$. Under Markov approximation, two-time correlation functions of these input noise operators are given by
	\begin{eqnarray}
		\langle a_{\rm 1,in}^\dag(t) a_{\rm 1,in}(t^\prime)\rangle\!&\!=\!&\!n_{\rm 1, th}\delta (t-t^\prime),\notag\\
	    \langle a_{\rm 1,in}(t) a_{\rm 1,in}^\dag(t^\prime)\rangle\!&\!=\!&\!(n_{\rm 1, th}+1)\delta (t-t^\prime),\notag\\
		\langle a_{\rm 2,in}^\dag(t) a_{\rm 2,in}(t^\prime)\rangle\!&\!=\!&\!n_{\rm 2,th}\delta (t-t^\prime),\notag\\
	\langle a_{\rm 2,in}(t) a_{\rm 2, in}^\dag(t^\prime)\rangle\!&\!=\!&\!(n_{\rm 2, th}+1)\delta (t-t^\prime),\\
	\langle b_{\rm in}^\dag(t) b_{\rm in}(t^\prime)\rangle\!&\!=\!&\!n_{\rm m,th}\delta (t-t^\prime),\notag\\
	\langle b_{\rm in}(t) b_{\rm in}^\dag(t^\prime)\rangle\!&\!=\!&\!(n_{\rm m, th}+1)\delta (t-t^\prime).\notag
	\end{eqnarray}
	Here $n_{\rm s, th}=(e^{\hbar\omega_s/k_B T_s}-1)^{-1}$ with $s=1,2,m$, is the mean photon or phonon number in the thermal bath, where $k_B$ is the Boltzmann constant and $T_s$ is the bath temperature. For zero temperature, i.e., $n_{\rm s, th}=0$. Now we write each operator of the system as its steady-state value plus the corresponding fluctuation, i.e., $a_{1(2)}\rightarrow \alpha_{1(2)} +a_{1(2)}$ and $b\rightarrow \beta+ b$. Substituting these transformations into Eq.~(\ref{s1}) and neglecting the higher-order fluctuation terms ensured by the strong laser fields, i.e., $\abs{\alpha_1},~\abs{\alpha_2}\gg1$, the quantum Langevin equations for the fluctuation operators can be linearized as
\begin{align}\label{s3}
	\dot{a}_1=&-(i\omega_{\rm 1,eff} +\kappa_1)a_1-iG_1 (b^\dag+b)+\sqrt{2\kappa_1}a_{\rm 1,in},\nonumber\\
	\dot{a}_2=&-(i\omega_{\rm 2,eff}+\kappa_2)a_1-iG_2 (b^\dag+b)+\sqrt{2\kappa_2}a_{\rm 2,in},\\
	\dot{b}=&-(i\omega_m+\gamma_m)b-i(G_1^* a_1+G_2^* a_2^\dag+{\rm H.c.})+\sqrt{2\gamma_m}b_{\rm in},\notag
\end{align}
where $\omega_{\rm 1(2),eff}=\delta_{1}\mp g_{1(2)}(\beta^*+\beta)$ is the effective frequency of cavity $1~(2)$ induced by the displacement of the MR. $G_{1(2)}=\mp g_{1(2)}\alpha_{1(2)}$ is the enhanced optomechanical coupling strength, which can be tuned by changing the powers of the laser fields. {By rewriting the equations of motion in Eq.~(\ref{s3}) as $\dot{a}_1=-i[a_1,H_{\rm eff}]+\sqrt{2\kappa_1}a_{\rm 1,in}$, $\dot{a}_2=-i[a_2,H_{\rm eff}]+\sqrt{2\kappa_2}a_{\rm 2,in}$, and $\dot{b}=-i[b,H_{\rm eff}]+\sqrt{2\gamma_m}b_{\rm in}$, we obtain the effective non-Hermitian Hamiltonian}
\begin{align}
	H_{\rm eff}=& (\omega_{\rm 1,eff}-i\kappa_1) a_1^\dag a_1+(\omega_{\rm 2,eff}-i\kappa_2) a_2^\dag a_2\nonumber\\&+(\omega_m-i\gamma_m)
	 b^\dag b+[(G_1 a_1^\dag+ G_2 a_2^\dag)b+{\rm H.c.}]\nonumber\\
	 &+[(G_1 a_1^\dag+ G_2 a_2^\dag)b^\dag+{\rm H.c.}],\label{q3}
\end{align}
 When $\delta_1,~\delta_2>0$, two cavities are in the red-detuning regime, where the rotating-wave approximation is allowed, i.e., fast oscillating terms in Eq.~(\ref{q3}) are neglected. Thus, Eq.~(\ref{q3}) reduces to
\begin{align}
H_{\rm eff}=& (\omega_{\rm 1,eff}-i\kappa_1) a_1^\dag a_1+(\omega_{\rm 2,eff}-i\kappa_2) a_2^\dag a_2\nonumber\\&+(\omega_m-i\gamma_m)
b^\dag b+[(G_1 a_1^\dag+ G_2 a_2^\dag)b+{\rm H.c.}],\label{q4}
\end{align}
which is the typically ideal Hamiltonian for the sideband cooling and energy conversion. For simplicity, $G_1$ and $G_2$ are assumed to be real hereafter, which can be achieved by tuning two laser fields.
\newcommand{\tabincell}[2]{}
\renewcommand\tabcolsep{55.0pt}
\begin{table*}[!t]
	\centering
	\scriptsize
	\caption{The most important symbols and some formulas }
	\label{tab:notations}
	\begin{tabular}{cc}
		\\[-1mm]
		\hline
		\hline\\[-2mm]
		{\bf \small Symbol}&\qquad {\bf\small Meaning}\\
		\hline
		\vspace{1mm}\\[-3mm]
		$\omega_{1(2)}$  & {Resonance frequency of cavity 1 (2)}\\
		\vspace{1mm}
		$\omega_{m}$  & {Resonance frequency of the MR}\\
		\vspace{1mm}
		$g_{1(2)}$  & Single-photon optomechanical coupling strength\\
		\vspace{1mm}
		{$\alpha_{1(2)}$}  & {The steady-state value of cavity 1(2)}\\
		\vspace{1mm}
		{$\beta$}  & {The steady-state value of the MR}\\
		\vspace{1mm}
		$G_{1(2)}$  & {$G_{1(2)}=\mp g_{1(2)}\alpha_{1(2)}$}, enhanced optomechanical coupling strength\\
		\vspace{1mm}
		$\Omega_{1(2)}$  & The amplitude of the drive field in cavity 1 (2)\\
		\vspace{1mm}
		{$\omega_0$}  & {The frequency of the drive fields in cavities 1 and 2}\\
		\vspace{1mm}
		$\delta_{1(2)}$  & {$\delta_{1(2)}=\omega_{1(2)}-\omega_0$},~detuning of cavity 1 (2) from the drive field\\
		\vspace{1mm}
		$\omega_{\rm 1(2),eff}$  & {$\omega_{\rm 1(2),eff}=\delta_{1}\mp g_{1(2)}(\beta^*+\beta)$, the effective frequency of the cavity 1(2)}\\
		\vspace{1mm}
		$\Delta_{1(2)}$  & {$\Delta_{1(2)}=$} $\omega_{\rm 1(2),eff}-\omega_m$, frequency detuning of the cavity from the MR \\
		\vspace{1mm}
		$\kappa_{1(2)}$  & The decay rate of cavity 1 (2)\\
		\vspace{1mm}
		$\gamma_m$  & The decay rate of the mechanical resonator\\
		\vspace{1mm}
		$\eta$  &  {$\eta=\kappa_1/\kappa_2$, the ratio of decay rates $\kappa_1$ and $\kappa_2$}\\
		\vspace{1mm}
		$\lambda$  & {$\lambda=G_2/G_1$, the ratio of coupling strengths $G_1$ and $G_2$}\\
		\vspace{1mm}
		$\Delta$  & The discriminant of the characteristic equation\\
		\vspace{1mm}
		$\Omega_{0,\pm}$  & Eigenvalues {of the pseudo-Hermitian system}\\
		\vspace{1mm}
		MR  & Mechanical resonator\\
		\vspace{1mm}
		COM  & Cavity optomechanical\\
		\vspace{1mm}
		EP2(3)  & The second-(third-) order exceptional point\\
		\\
		\hline
		\hline
	\end{tabular}
\end{table*}

\section{Pseudo-Hermitian condition}

In matrix form, Eq.~(\ref{q4}) can be expressed as
\begin{equation}\label{q9}
H_{\rm eff}=\left(
\begin{array}{ccc}
\omega_{\rm 1,eff}-i\kappa_1 & 0 & G_1\\
0 & \omega_{\rm 2,eff}-i\kappa_2 & G_2\\
G_1 &  G_2  & \omega_m-i\gamma_m
\end{array}
\right).
\end{equation}
For this considered Hamiltonian, three eigenvalues can be predicted. In particular, when three eigenvalues are all real or one of the three eigenvalues is real and the other two are a complex-conjugate pair, the system governed by the Hamiltonian in Eq.~(\ref{q9}) is a pseudo-Hermitian system. According to the energy-spectrum properties of pseudo-Hermitian systems, the solutions of the characteristic equation $\abs{H_{\rm eff}-\Omega I}=0$ are the same as that of $\abs{H_{\rm eff}^*-\Omega I}=0$, where $I$ is an identity matrix, and $\Omega$ is the eigenvalue of the Hamiltonian $H_{\rm eff}$. These two equations give rise to the pseudo-Hermitian conditions of the Hamiltonian~(\ref{q9}) as
\begin{align}\label{q11}
\gamma_m+(1+\eta)\kappa_2=&0,\nonumber\\
\Delta_1\eta+\Delta_2=&0,\nonumber\\
(1+\eta \lambda^2)G_1^2\kappa_2+\gamma_m(\Delta_1^2+\kappa_2^2)\eta=&0,
\end{align}
where $\Delta_{1(2)}=\omega_{\rm 1(2),eff}-\omega_m$ is the detuning of the effective cavity frequency from the resonator. $\eta=\kappa_1/\kappa_2$ is introduced to characterize the symmetry ($\eta=1$) and asymmetry ($\eta\neq 1$) between $\kappa_1$ and $\kappa_2$, and $\lambda=G_2/G_1$ is the relative strength of $G_2$ and $G_1$. Equation~(\ref{q11}) shows that the proposed system is pseudo-Hermitian only when these three conditions are simultaneously satisfied. The first condition requires that the total decay rates of the system are zero, which shows that the gain effect must be introduced. The second condition reveals that the ratio of $\Delta_2/\Delta_1$ must match the ratio of $\kappa_1/\kappa_2$, which can be realized in our considered system due to the tunable $\Delta_1$ and $\Delta_2$. The third condition in Eq.~(\ref{q11}) indicates that the enhanced coupling strengths $G_1$ and $G_2$ are bounded by other system parameters, which can also be achieved here owing to the controllable parameters $G_1$, $G_2$, $\Delta_1$, and $\Delta_2$.

With the conditions in Eq.~(\ref{q11}) (i.e., the system is pseudo-Hermitian), the characteristic equation $\abs{H_{\rm eff}-\Omega I}=0$ can be specifically written as
\begin{align}
	(\Omega-\omega_m)^3+c_2 (\Omega-\omega_m)^2+c_1 (\Omega-\omega_m)+c_0=0,\label{q12}
\end{align}
where
\begin{align}\label{q13}
    c_0=&(\lambda^2-\eta )G_1^2\Delta_1+\gamma_m(1-\eta^2)\Delta_1\kappa_2,\nonumber\\
	c_1=&\gamma_m^2-\eta(\Delta_1^2+\kappa_2^2)-(1+\lambda^2)G_1^2,\nonumber\\
	c_2=&(\eta-1)\Delta_1.	
\end{align}
According to Cardano's formula method~\cite{kORN-1968}, the solutions of this characteristic equation can be determined by the discriminant
\begin{align}
	\Delta=B^2-4AC\label{s11},
\end{align}
with
\begin{align}
	A=&c_2^2-3c_1,\nonumber\\
	 B=&c_1c_2-9c_0,\nonumber\\
	  C=&c_1^2-3c_0c_2.
\end{align}
For $\Delta<0$, Eq.~(\ref{q12}) has three real roots. But when $\Delta>0$, only one real root survives, and the other two become complex conjugates. In the critical case, i.e., $\Delta=0$, three real roots coalesce to the same value,  $\Omega=\Omega_{\rm EP3}$, when $A=B=0$, corresponding to EP3. But when $A\neq0$ and $B\neq0$, only two real roots of Eq.~(\ref{q12}) coalesce to a certain value, $\Omega=\Omega_{\rm EP2}$, which is the typical EP2.

\section{EP3 in a pseudo-Hermitian COM system without $\mathcal{PT}$  symmetry}\label{IV}

From the first condition in Eq.~(\ref{q11}), the gain effect must be introduced to the proposed system to keep the gain-loss balance. For this, we here consider that the MR is active and two cavities are passive, i.e., $\gamma_m<0$ and $\kappa_1,\kappa_2>0$.  According to above analysis, conditions in Eq.~(\ref{q11}) convert the non-Hermitian Hamiltonian in Eq.~(\ref{q9}) into a pseudo-Hermitian Hamiltonian. Such a Hamiltonian does not have the $\mathcal{PT}$ symmetry, but it can be used to predict the EPs such as EP3 and EP2. To prove this,  in Fig.~\ref{fig2}(a) we plot the phase diagram determined by the sign of the discriminant in Eq.~(\ref{s11}), where the yellow (white) area denotes $\Delta<0~(\Delta>0)$ and the red dashed line represents $\Delta=0$. {Note that the region circled by a green curve must be taken out, where the parameter value of $\Delta_2^2<0$ is inaccessible in the realistic COM system.}
Also $A=0$ and $B=0$ are shown by the blue and black dashed lines, respectively, in Fig.~\ref{fig2}(a). According to Cardano formula~\cite{kORN-1968}, the crossing points produced by the red, blue, and black lines denote EP3s, and the points determined by only the red line are EP2s. Obviously, EP3s (denoted by the green dot) can be predicted by tuning the couplings $G_1$ and $G_2$.

Below we analytically derive the critical parameters for observation of EP3. We assume that EP3 is predicted at $\Omega=\Omega_{\rm EP3}$, which means $(\Omega-\Omega_{\rm EP3})^3=0$.  Comparing this equation with Eq.~(\ref{q12}), we have
\begin{align}\label{q22}
	-3(\Omega_{\rm EP3}-\omega_m)=&c_2,\nonumber\\
	3(\Omega_{\rm EP3}-\omega_m)^2=&c_1,\nonumber\\
	-(\Omega_{\rm EP3}-\omega_m)^3=&c_0.
\end{align}
The first equation in Eq.~(\ref{q22}) directly gives
\begin{align}\label{q23}
	\Omega_{\rm EP3}=\frac{1}{3}(1-\eta)\Delta_{\rm 1, EP3}+\omega_m.
\end{align}
Using the second equation in Eq.~(\ref{q22}) and the third condition in Eq.~(\ref{q11}), the coupling strength $G_1$ between cavity 1 and the MR at EP3 is
\begin{align}\label{q25}
	G_{\rm 1,EP3}=&2\kappa_2\Bigg[\frac{3(1+\lambda_{\rm EP3}^2)}{1+\eta+\eta^2}
	+\frac{1+\eta\lambda_{\rm EP3}^2}{(1+\eta)\eta}\Bigg]^{-1/2},
\end{align}
where
\begin{align}\label{q26}
	\lambda_{\rm EP3}=\Bigg[\frac{1+2\eta}{(2+\eta)\eta}\Bigg]^{3/2}
\end{align}
is given by the third equation in Eq.~(\ref{q22}). This indicates that the coupling strength $G_2$ between cavity 2 and the MR at EP3 is
\begin{align}
	G_{\rm 2,EP3}=	\lambda_{\rm EP3}G_{\rm 1,EP3},\label{s26}
\end{align}
{When the optomechanical coupling strengths $G_1$ and $G_2$ are tuned to satisfy Eq.~(\ref{s26}), one EP2 can also be predicted [see the green point formed by the red and gray curves in Fig.~\ref{fig2}(b)], at which $\Delta=0$ but $A\neq 0$ and $B\neq 0$. When Eqs.~(\ref{q25})-(\ref{s26}) are kept, the parameter $\Delta_1$ at EP3 accordingly becomes}
\begin{align}\label{q27}
\Delta_{\rm 1,EP3}^2=\frac{1+\eta\lambda_{\rm EP3}^2}{\eta(1+\eta)}G_{\rm 1,EP3}^2-\kappa_2^2,
\end{align}
which gives rise to the minimal $G_{\rm 1,EP3}$ for predicting EP3,
\begin{align}
G_{\rm 1,EP3}^{\rm min}=\sqrt{\frac{\eta(1+\eta)}{1+\eta\lambda_{\rm EP3}^2}}\kappa_2.\label{q28}
\end{align}
When conditions in Eqs.~(\ref{q25})-(\ref{q28}) are simultaneously satisfied, the pseudo-Hermitian COM system can be used to predict EP3.

\begin{figure}
	\center
	\includegraphics[scale=0.6]{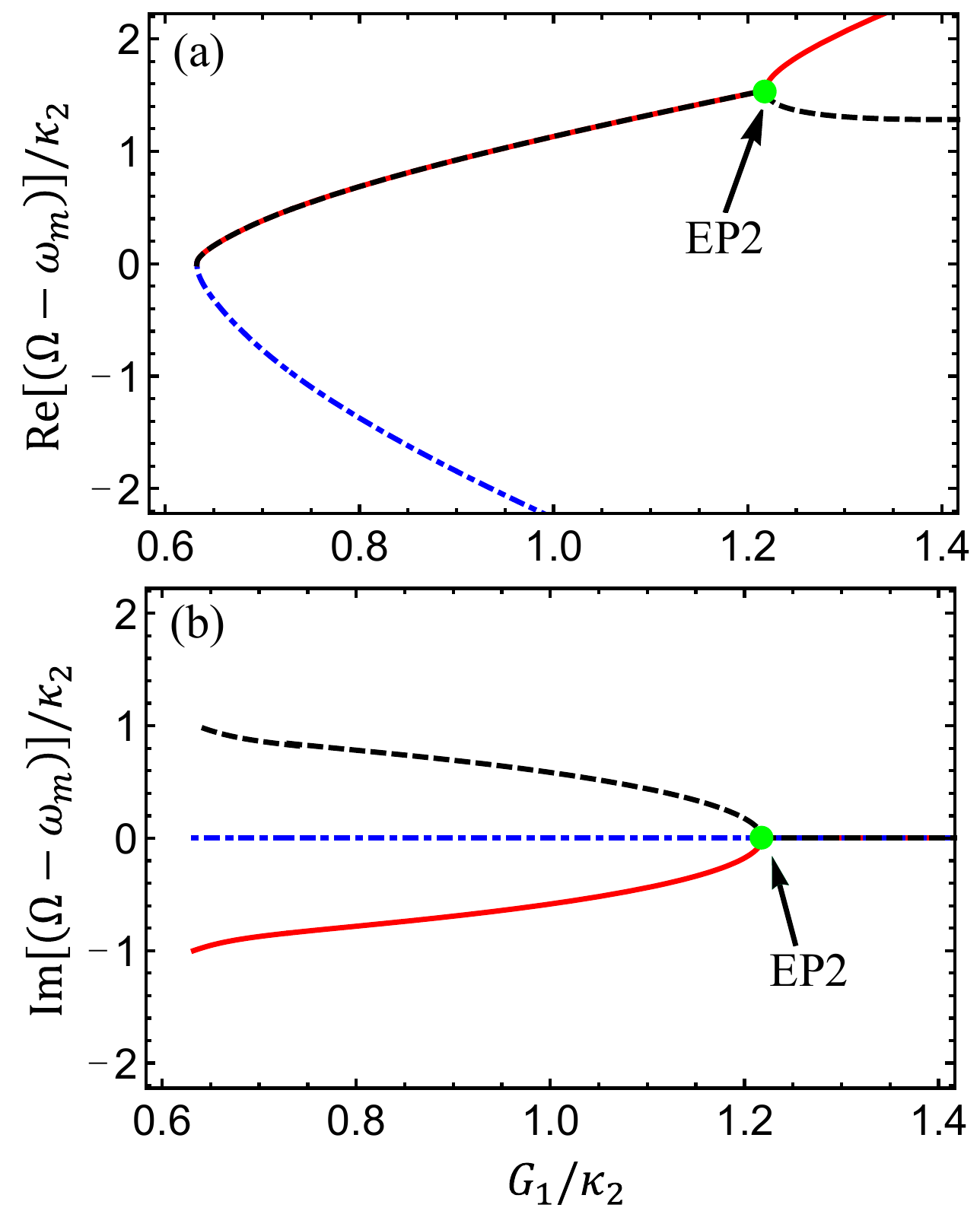}
	\caption{(a) The real and (b) imaginary parts of the eigenvalues given by Eq.~(\ref{q12}) with $\gamma_m=-(1+\eta)\kappa_2<0$ vs the normalized parameter $G_1/\kappa_2$. Here $\lambda=2$ is chosen to break the parameter condition at EP3 in Eq.~(\ref{q26}) and $\eta=1$.}
	\label{fig4}
\end{figure}

\begin{figure}
	\center
	\includegraphics[scale=0.6]{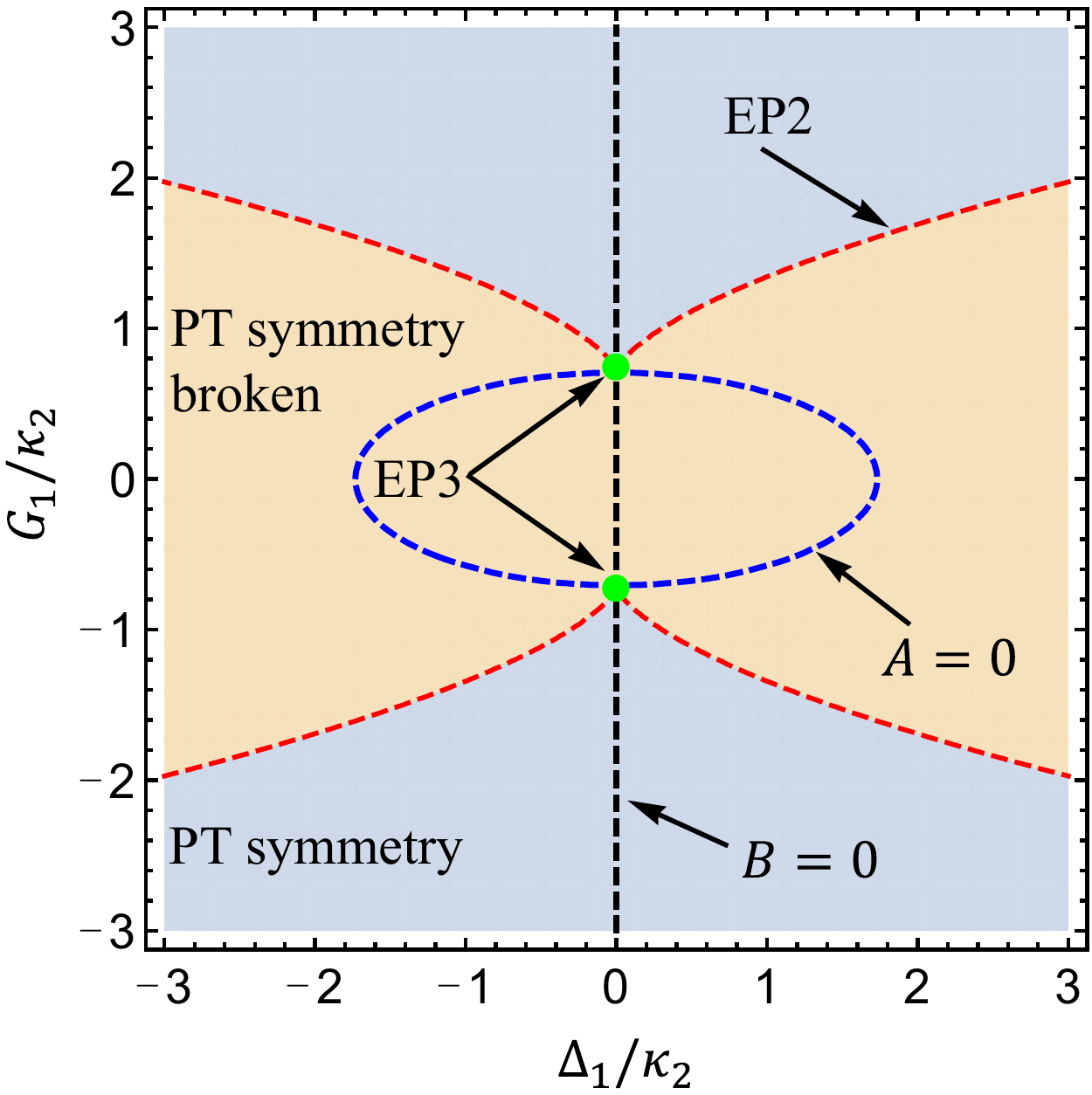}
	\caption{The phase diagram of the discriminant given by Eq.~(\ref{s11}) with $\gamma_m\ll\kappa_2=-\kappa_1$ vs the normalized parameters $G_1/\kappa_2$ and $\Delta_1/\kappa_2$, where $\kappa_1$ denotes the gain of cavity 1. Here $G_2=G_1$ is chosen.}
	\label{fig5}
\end{figure}

\subsection{$\eta=1$}
Now we further discuss the effect of the parameter $\eta$ on the EP of the proposed COM system. {For the simplest case, we consider two symmetric cavities with the identical decay rates, $\kappa_1=\kappa_2$,}
corresponding to $\eta=1$. In this symmetric case, the pseudo-Hermitian conditions in Eq.~(\ref{q11}) reduce to
\begin{align}
	\gamma_m=&-2\kappa_2,~~\Delta_2=-\Delta_1,\nonumber\\
	\Delta_1^2=&\frac{1}{2}(1+\lambda^2)G_1^2-\kappa_2^2.
\end{align}
Correspondingly, the coefficients in Eq.~(\ref{q13}) become
\begin{align}
	c_0=&c_2=0,\nonumber\\
	c_1=&4\kappa_2^2-\frac{3}{2}(1+\lambda^2)G_1^2,\label{q19}
\end{align}
and the discriminant is
\begin{align}
	\Delta=B^2-4AC=12 c_1^3.
\end{align}
When $\Delta=0$, $c_1=0$, which leads to $A=B=0$. This indicates that only EP3 can be predicted in the proposed COM system with two symmetric cavities.

More specifically, we substitute the coefficients in Eq.~(\ref{q19}) into Eq.~(\ref{q12}), and we have
\begin{align}
[(\Omega-\omega_m)^2+c_1] (\Omega-\omega_m)=0.\label{eq21}
\end{align}
This immediately gives rise to three roots,
\begin{align}
\Omega_0=&\omega_m,~~
\Omega_\pm=\omega_m\pm\sqrt{-c_1}.
\end{align}
Obviously, Eq.~(\ref{eq21}) has three different real roots for $c_1<0$ (i.e., $\Delta<0$). For $c_1>0$ (i.e., $\Delta>0$), Eq.~(\ref{eq21}) has one real root $\Omega_0=\omega_m$ and two complex roots $\Omega_\pm$. This indicates that the polariton mode with the eigenvalue $\Omega_0$ has a zero loss rate, and the upper (lower) polariton mode with the eigenvalue $\Omega_+$~($\Omega_-$) has a gain (loss) rate. In particular, three real roots coalesce into one when $c_1=0$, which is the EP3 of the proposed COM system. To show this prediction, we plot the real and imaginary parts of the eigenvalue $\Omega$ of the Hamiltonian (\ref{q9}) with $\eta=1$ as a function of the normalized parameter $G_1/\kappa_2$ in Figs.~\ref{fig3}(a) and \ref{fig3}(b), respectively. When $\kappa_2=G_{\rm 1,EP3}^{\rm min}\leq G_1<G_{\rm 1,EP3}$, $\Omega_0$ is real (see the red curve), and $\Omega_\pm$ are a complex-conjugate pair (see the black and blue curves). At $G_1=G_{\rm 1,EP3}$, three eigenvalues ($\Omega_0,~\Omega_\pm$) coalesce into $\Omega_{\rm EP3}=\omega_m$. By going to increasing $G_1$, the three eigenvalues are all real but bifurcate into three different values. At EP3, the corresponding parameters are
\begin{align}
\lambda_{\rm EP3}|_{\eta=1}=&1,\nonumber\\
	G_{\rm 1,EP3}|_{\eta=1}=&\frac{2}{\sqrt{3}}\kappa_2,\nonumber\\
	\Delta_{\rm 1,EP3}|_{\eta=1}=&-\frac{1}{\sqrt{3}}\kappa_2.
\end{align}

\begin{figure}
	\center
	\includegraphics[scale=0.34]{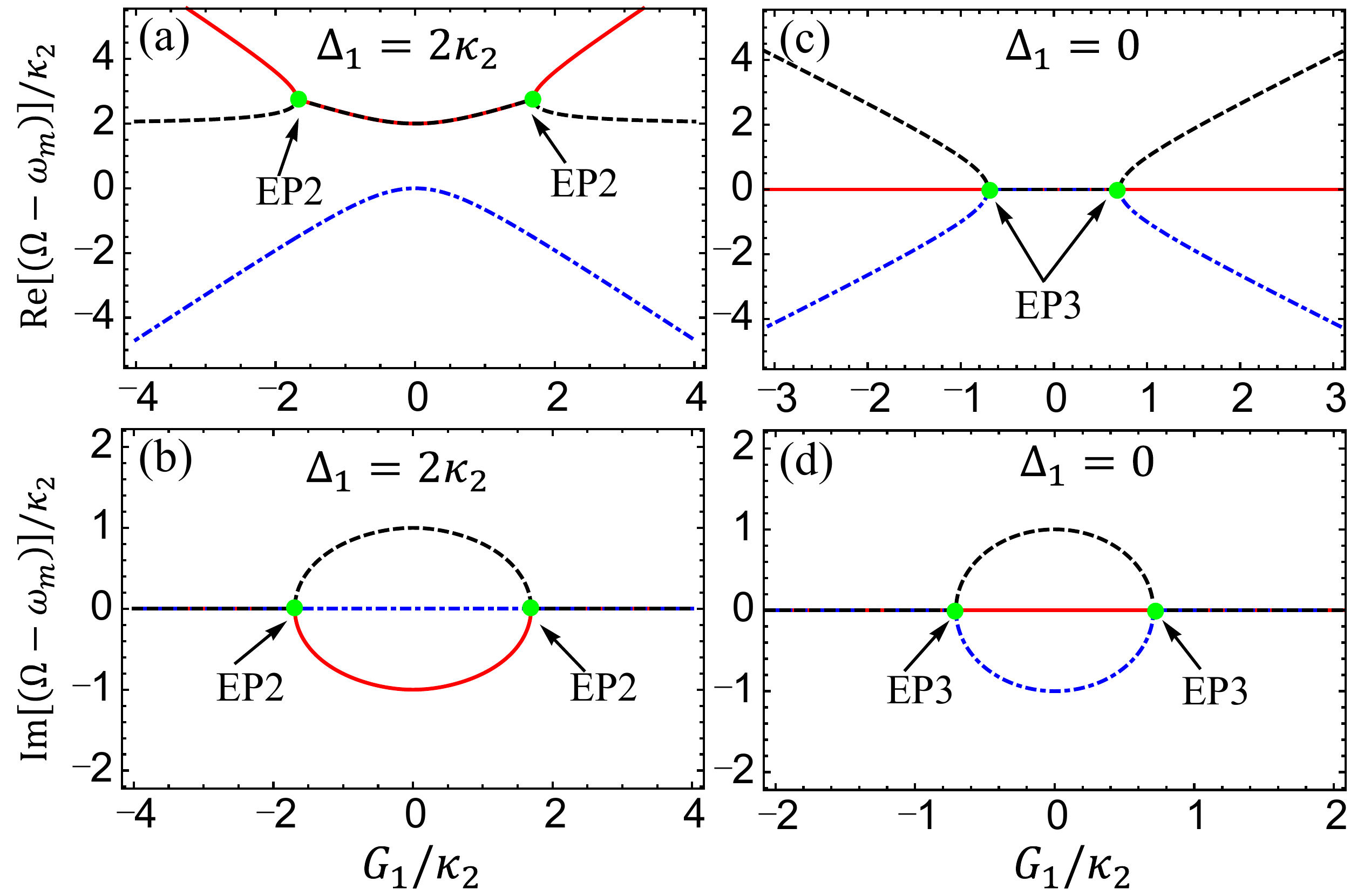}
	\caption{The real and imaginary parts of the eigenvalues given by Eq.~(\ref{qq}) with $\gamma_m\ll\kappa_2=-\kappa_1$ vs the normalized parameter $G_1/\kappa_2$ for different $\Delta_1$. In (a) and (b), $\Delta_1=2\kappa_2$ for predicting EP2. In (c) and (d), $\Delta_1=0$ for predicting EP3.}
	\label{fig6}
\end{figure}

\begin{figure}
	\center
	\includegraphics[scale=0.34]{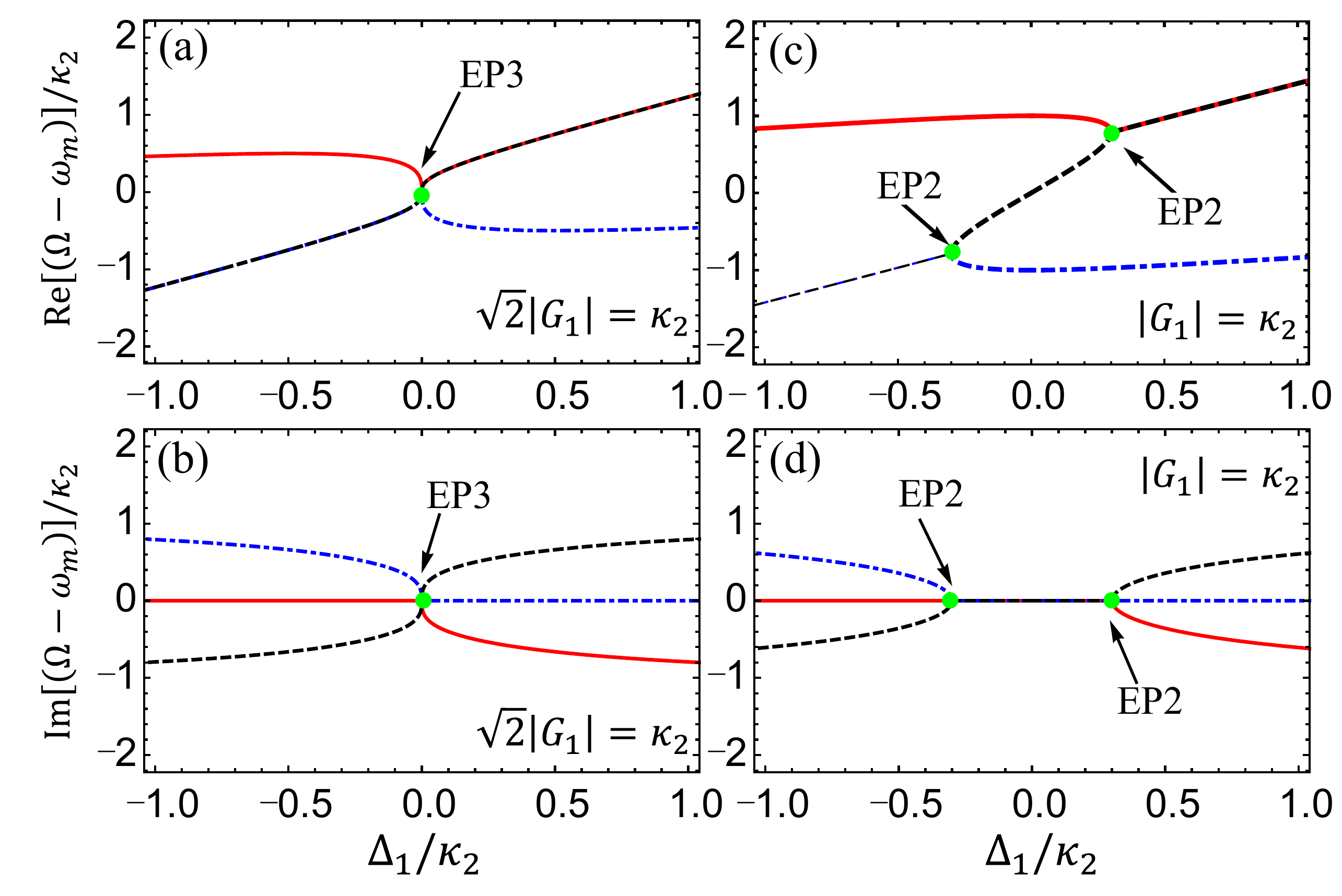}
	\caption{The real and imaginary parts of the eigenvalues given by Eq.~(\ref{qq}) with $\gamma_m\ll\kappa_2=-\kappa_1$ vs the normalized parameter $\Delta_1/\kappa_2$ for different $G_1$. In (a) and (b), $\abs{G_1}=\kappa_2/\sqrt{2}$ for predicting EP3. In (c) and (d), $\abs{G_1}=\kappa_2$ for predicting EP2.}
	\label{fig7}
\end{figure}

\subsection{$\eta\neq 1$}

In practice, fabricating two cavities with the same decay rates is hard due to the experimental errors. Hence, studying the EP in the proposed COM system using two cavities with different decay rates is full of realistic significance. For this, we here consider two asymmetric cavities with $\kappa_1\neq\kappa_2$, {i.e.,} $\eta\neq1$. As an example, we take $\eta=2$, which leads to $\lambda_{\rm EP3}=0.494$ and $G_{\rm 1,EP3}^{\rm min}=2$. As the analytical solution of Eq.~(\ref{q12}) is tedious, we numberically plot the real and imaginary parts of the eigenvalue $\Omega$ of the Hamiltonian (\ref{q9}) with $\eta=2$ as a function of the normalized parameter $G_1/\kappa_2$ in Figs.~\ref{fig3}(c) and \ref{fig3}(d), respectively. It is not difficult to find one of the eigenvalues ($\Omega_0$) is real for arbitrary $G_1$ (see the black curve). The other two eigenvalues ($\Omega_\pm$) are a complex-conjugate pair (see the blue and red curves) when $2\kappa_2\leq G_1<G_{\rm 1,EP3}=2.263\kappa_2$. At $G_1=G_{\rm 1,EP3}=2.263\kappa_2$, three eigenvalues coalesce to $\Omega_{\rm EP3}=\omega_m-0.173\kappa_2$. By sequentially increasing $G_1$ to $G_{\rm 1,EP3}\leq G_1<G_{\rm 1, EP2}=2.4\kappa_2$, $\Omega_{\pm}$ become complex again (see the blue and red curves). At $G_1=G_{\rm 1, EP2}$, these two eigenvalues coalesce into $\Omega_{\rm EP2}=\omega_m-0.842\kappa_2$. When $G_1>G_{\rm 1, EP2}$, $\Omega_{\pm}$ are real but bifurcate into two different values. For other values of $\eta$ ($\neq1$), the similar results can also be numerically obtained, which can easily be demonstrated.

We note that EP2 is predicted in Figs.~\ref{fig3}(c) and \ref{fig3}(d) when Eq.~(\ref{q26}) is satisfied, which is one of the conditions for observing EP3. Actually, this condition is not a necessary condition for predicting EP2. In Fig.~\ref{fig4}, we numerically plot the real and imaginary parts of the eigenvalue $\Omega$ of the Hamiltonian (\ref{q9}) with $\eta=1$ as a function of the normalized parameter $G_1/\kappa_2$ when $\lambda=2\neq \lambda_{\rm EP3}$ (i.e., $G_2=2 G_1$). We find EP2 still appears at $G_1/\kappa_2=1.22$. For the case of $\eta=2$, we also numerically check it, and the same result is obtained.

\section{EP3 in a pseudo-Hermitian COM system with $\mathcal{PT}$ symmetry}

As a special case of the pseudo-Hermitian systems, the $\mathcal{PT}$ symmetric system with EPs exhibits amazing characteristics and has wide applications in quantum information science~\cite{Bergholtz-2021,Wiersig-2020,Feng-2017,El-Ganainy-2018}. Next, we investigate the EPs in the pseudo-Hermitian COM system with $\mathcal{PT}$ symmetry.
We here consider the case that one cavity (cavity 2) and the mechanical resonator are passive and the other cavity (cavity 1) is active, i.e., $\gamma_m,~\kappa_2>0$ and $\kappa_1<0$. For experimental COM systems, $\gamma_m\ll\kappa_2$ in general, so we can safely ignore the effect of $\gamma_m$ by assuming $\gamma_m=0$. To meet the first condition in Eq.~(\ref{q11}), $\eta=-1$ is taken; that is, $\kappa_1=-\kappa_2$. This leads to $\Delta_2=\Delta_1$, and the third condition in Eq.~(\ref{q11}) is always valid for arbitrary parameters. With these parameters, the Hamiltonian of the proposed system given by Eq.~(\ref{q9}) reduces to

\begin{equation}\label{qq}
H_{\rm \mathcal{PT}}=\left(
\begin{array}{ccc}
\omega_{\rm 1,eff}+i\kappa_2 & 0 & G_1\\
0 & \omega_{\rm 2,eff}-i\kappa_2 & G_2\\
G_1 &  G_2  & \omega_m
\end{array}
\right).
\end{equation}
Obviously, this Hamiltonian is $\mathcal{PT}$ symmetric due to the invariant by simultaneously performing the following operations: $a_1\leftrightarrow a_2$ (corresponding to the $\mathcal
{P}$ operation) and $i\leftrightarrow-i,~a_{1(2)}\leftrightarrow-a_{1(2)},~b\leftrightarrow-b$ (corresponding to the $\mathcal{T}$ operation). To judge EPs in such a system, we plot the phase diagram of the discriminant given by Eq.~(\ref{s11}) in Fig.~\ref{fig5}, where $\lambda=1$ (i.e., $G_1=G_2$) is chosen. $\Delta<0$ indicates the system is in the $\mathcal{PT}$ symmetric phase (see the blue area), where all three eigenvalues for the Hamiltonian (\ref{q9}) are real. $\Delta>0$ means the system is in the $\mathcal{PT}$-symmetry-broken phase (see the yellow area), where only one real eigenvalue survives and the other two become complex conjugates for the Hamiltonian (\ref{q9}). Critically, $\Delta=0$ is denoted by the red dashed curve. Also, $A=0$ and $B=0$ are plotted, respectively, by blue and black dashed lines. Clearly, the three curves (red, blue, and black) give two crossing points, which correspond to two EP3s. EP2 can be predicted along the red curve, where $A\neq 0$ and $B\neq0$.

More specifically, we plot the real and imaginary parts of the eigenvalues versus the normalized parameter $G_1/\kappa_2$ with $\Delta_1=2\kappa_2$ and $\Delta_1=0$ in Fig.~\ref{fig6}. From Figs.~\ref{fig6}(a) and \ref{fig6}(b), we find that the system can have only EP2. By changing $G_1$, the three eigenvalues can have different characteristics. The eigenvalue denoted by the blue curve is real for arbitrary $G_1$,  while the other two eigenvalues (denoted by the black and red curves) are real only when $\abs{G_1}>G_{\rm 1, EP2}=1.692\kappa_2$. For $\abs{G_1}\leq G_{\rm 1, EP2}$, these two eigenvalues become a complex-conjugate pair, where the equality means two eigenvalues coalesce into one real value, $\Omega_\pm=\Omega_{\rm EP2}=2.755\kappa_2$.  When $\Delta_1=0$ in Figs.~\ref{fig6}(c) and \ref{fig6}(d), the eigenvalue $\Omega_0$ is real for arbitrary $G_1$, while $\Omega_\pm$ are real for only $\abs{G_1}>G_{\rm 1, EP3}$. When $\abs{G_1}<G_{\rm 1, EP3}$, $\Omega_\pm$ are a complex-conjugate pair. At the critical value $G_1=G_{\rm 1, EP3}$, three eigenvalues coalesce into $\Omega=\Omega_{\rm EP3}=\omega_m$, corresponding to EP3.

In Fig.~\ref{fig7}, we plot the real and imaginary parts of the eigenvalues versus the normalized parameter $\Delta_1/\kappa_2$ with $\sqrt{2}\abs{G_1}=\kappa_2$ and $\abs{G_1}=\kappa_2$. For $\abs{G_1}=\kappa_2/\sqrt{2}$ in Figs~\ref{fig7}(a) and \ref{fig7}(b), we find EP3 can be predicted at $\Delta_{\rm 1,EP3}=0$, where three eigenvalues coalesce to a certain value. When $\Delta_1<\Delta_{\rm 1,EP3}$, the eigenvalue marked in red is real, and the eigenvalues marked in blue and black are a complex-conjugate pair. However, the eigenvalue marked in blue becomes real, and the other two eigenvalues are a complex-conjugate pair for $\Delta_1>\Delta_{\rm 1,EP3}$. When $\abs{G_1}=\kappa_2$, the system can exhibit EP2, as shown in Figs.~\ref{fig7}(c) and \ref{fig7}(d). For $\Delta_1<\Delta_{\rm 1,EP2}^{(-)}=-0.3$, the eigenvalue marked in red is real, and the other two eigenvalues, marked in blue and black, are a complex-conjugate pair. But for $\Delta_1>\Delta_{\rm 1,EP2}^{(+)}=0.3$, the eigenvalue denoted by the blue curve is real, and the eigenvalues plotted in black and red become a complex-conjugate pair. When $\Delta_{\rm 1,EP2}^{(-)}<\Delta_1<\Delta_{\rm 1,EP2}^{(+)}$, the three eigenvalues are all real but have different values. At the point $\Delta_1=\Delta_{\rm 1,EP2}^{(-)}$,the two eigenvalues marked in black and blue coalesce to the value $\Omega_{\rm EP2}=\omega_m-0.8\kappa_2$, which is EP2. In addition, we find eigenvalues marked in black and red coalesce to the value $\Omega_{\rm EP2}=\omega_m+0.8\kappa_2$ at the point $\Delta_1=\Delta_{\rm 1,EP2}^{(+)}$, which is also EP2.

\section{Conclusion}

In summary, we have proposed a pseudo-Hermitian COM system consisting of two cavities coupled to a common MR via radiation pressure for predicting EP3s. We showed that under certain conditions the non-Hermitian COM system can be equivalent to a pseudo-Hermitian system without $\mathcal{PT}$ symmetry hosting both EP3 and EP2 in the parameter space when mechanical gain is taken into account. In this case, only EP3 or EP2 can be predicted when two symmetric cavities are considered. But for asymmetric cavities, EP3 and EP2 can be transformed into each other by tuning the COM coupling strength. In another case, the non-Hermitian COM system can be reduced to a pseudo-Hermitian system with  $\mathcal{PT}$ symmetry when one of the cavities with gain is considered. Such a $\mathcal{PT}$-symmetric Hamiltonian can be used to predict EP3 and EP2 in parameter space. We further specifically considered the impacts of the system parameters such as the optomechanical coupling strength and frequency detuning on them. Our proposal may provide a path to study physical phenomena around higher-order EPs in COM systems.

\section*{ACKNOWLEDGMENTS}

This paper is supported by the National Natural Science Foundation of China (Grant No.~11804074), and the Postdoctoral Science Foundation of China (Grant No. 2020M671687).

\appendix

\section{eigenvectors of the proposed system without $\mathcal{PT}$ symmetry}\label{appendix-A}

In the main text, only the eigenvalues of the effective Hamiltonian given by Eq.~(\ref{q9}) are discussed. As is well known, eigenvalues and eigenvectors at EPs  are required to coalesce simultaneously. For this, we further show that the eigenvectors of the effective Hamiltonian~(\ref{q9}) coalesce at EP3 and EP2 in this appendix.

We first study the eigenvectors of the pseudo-Hermitian COM system without $\mathcal{PT}$ symmetry. Two cases are considered: the symmetric case ($\eta=1$) and the asymmetric case ($\eta\neq1$). For simplicity, we assume $\omega_m=0$ in this appendix, which is equivalent to rotating the system governed by Eq.~(\ref{q4}) at the mechanical frequency $\omega_m$.

\subsection{The case of $\eta=1$}

For the symmetric case of $\eta=1$, the effective Hamiltonian in Eq.~(\ref{q9}) at EP3 can be written as
\begin{equation}\label{a1}
	H_{\rm EP3}^{(\eta=1)}=\left(
	\begin{array}{ccc}
	-1/\sqrt{3}-i & 0 & 2/\sqrt{3}\\
	0 & 1/\sqrt{3}-i & 2/\sqrt{3}\\
	2/\sqrt{3} &  2/\sqrt{3}  & 2i
	\end{array}
	\right)\kappa_2.
\end{equation}
The eigenvalues of $H_{\rm EP3}^{(\eta=1)}$ are $\Omega_\pm=\Omega_0=0$, and the corresponding eigenvectors are $|\Omega_\pm\rangle=|\Omega_0\rangle=(0.5 - 0.866025 i, -0.5 - 0.866025 i, 1)^T$.  This indicates that  eigenvectors and the corresponding three eigenvalues actually coalesce simultaneously at the EP3 shown in Figs.~\ref{fig3}(a) and \ref{fig3}(b) of the main text.

\subsection{The case of $\eta\neq1$}
For the asymmetric case of $\eta\neq 1$, we take $\eta=2$ as an example, which is consistent with the situation discussed in Figs.~\ref{fig3}(c) and \ref{fig3}(d) of the main text. At EP3, the effective Hamiltonian in Eq.~(\ref{q9}) of the main text becomes
\begin{equation}\label{a2}
H_{\rm EP3}^{(\eta=2)}=\left(
\begin{array}{ccc}
	0.520-2i & 0 & 2.263\\
	0 & -1.039-i & 1.118\\
	2.263 &  1.118  & 3i
\end{array}
\right)\kappa_2.
\end{equation} By diagonalizing this Hamiltonian, we find three eigenvalues coalesce into $\Omega_\pm=\Omega_0=-0.173\kappa_2$. Correspondingly, the three eigenvectors are $|\Omega_\pm\rangle=|\Omega_0\rangle=(0.632, 0.250 + 0.433i, -0.194 + 0.559i)^T$, indicating that the three eigenvectors actually coalesce at the EP3 where the three eigenvalues degenerate.

We next investigate the eigenvectors at EP2 shown in Figs.~\ref{fig3}(c) and \ref{fig3}(d). At this point, the effective Hamiltonian in Eq.~(\ref{q9}) reduces to
\begin{equation}\label{a3}
	H_{\rm EP2}^{(\eta=2)}=\left(
	\begin{array}{ccc}
	0.520-2i & 0 & 2.263\\
	0 & -1.039-i & 1.118\\
	2.263 &  1.118  & 3i
	\end{array}
	\right)\kappa_2.
	\end{equation}
For this Hamiltonian, its three eigenvalues are $\Omega_0=1.029\kappa_2$, $\Omega_+=\Omega_-=-0.842\kappa_2$, which shows two eigenvalues coalesce. The three corresponding three eigenvectors are $|\Omega_0\rangle=(0.730, 0.160 + 0.240i, 0.114 + 0.609i)^T$ and $|\Omega_+\rangle=|\Omega_-\rangle=(0.258 - 0.483i, 0.612 , 0.242 + 0.516i)^T$. This indicates two eigenvectors coalesce at the point where $\Omega_\pm$ degenerate.

\section{eigenvectors of the proposed system with $\mathcal{PT}$ symmetry}\label{appendix-B}

We now study the eigenvectors of the pseudo-Hermitian COM system with $\mathcal{PT}$ symmetry. In this situation, $\eta=-1$ (or $\kappa_1=-\kappa_2$), $\lambda=1$ (i.e., $G_2=G_1$) and $\Delta_2=\Delta_1$ are required, and the system Hamiltonian is governed by Eq.~(\ref{qq}). Here we  take only Fig.~\ref{fig6} as an example for investigating the eigenvectors of the Hamiltonian in Eq.~(\ref{qq}). In Fig.~\ref{fig6}, $\Delta_1=2\kappa_2$ and $\Delta_1=0$ are discussed. When $\Delta_1=2\kappa_2$, Eq.~(\ref{qq}) becomes
\begin{equation}\label{b1}
	H_{\rm EP2}^{(\Delta_1=2\kappa_2)}=\left(
	\begin{array}{ccc}
	2+i & 0 & 1.692\\
	0 & 2-i & 1.692\\
	1.692 &  1.692  & 0
	\end{array}
	\right)\kappa_2.
	\end{equation}
The eigenvalues are $\Omega_0=-1.510\kappa_2$ and $\Omega_+=\Omega_-=2.755\kappa_2$. The corresponding eigenvectors are $|\Omega_0\rangle=(-0.373 + 0.106i, -0.373 - 0.106i, 0.836)$, and $|\Omega_+\rangle=|\Omega_-\rangle=(0.626, -0.172 - 0.602i, 0.279- 0.370i)$. This shows that two eigenvalues and the corresponding eigenvectors of the $\mathcal{PT}$ symmetric Hamiltonian in Eq.~(\ref{qq}) simultaneously coalesce at EP2 discussed in Figs.~\ref{fig6}(a) and \ref{fig6}(b) of the main text.

For $\Delta_1=0$, Eq.~(\ref{qq}) can be written as
\begin{equation}\label{b2}
	H_{\rm EP3}^{(\Delta_1=0)}=\left(
	\begin{array}{ccc}
	i & 0 & \sqrt{2}/2\\
	0 & -i & \sqrt{2}/2\\
	\sqrt{2}/2 &  \sqrt{2}/2  & 0
	\end{array}
	\right)\kappa_2.
	\end{equation}
Such a Hamiltonian has three degenerate eigenvalues, $\Omega_\pm=\Omega_0=0$. The corresponding eigenvectors are $|\Omega_\pm\rangle=|\Omega_0\rangle=(0.707i, - 0.707i, 1)^T$. Thus, the three eigenvalues and eigenvectors coalesce simultaneously at EP3.

\section{Cubic root dependence of eigenvalues near EP3}\label{appendix-C}

In this appendix, we consider the case in which the pseudo-Hermitian COM system is disturbed by a small parameter $\epsilon$  near the EPs, to predict the behavior of the eigenvalues with this perturbation. Without loss of generality, we assume the perturbation is applied on the frequency of cavity 1. We start from the general non-Hermitian Hamiltonian given by Eq.~(\ref{q9}). Thus, the perturbation Hamiltonian of the system can be written as
\begin{equation}\label{c1}
\tilde{H}=\left(
\begin{array}{ccc}
\Delta_1-i\kappa_1+\epsilon & 0 & G_1\\
0 & \Delta_2-i\kappa_1 & G_2\\
G_1 &  G_2  & -i\gamma_m
\end{array}
\right).
\end{equation}
The corresponding characteristic equation $|\tilde{H}-\tilde{\Omega}I|=0$ can be specifically expressed as
\begin{align}\label{c2}
	\tilde{\Omega}^3+C_2\tilde{\Omega}^2+C_1\tilde{\Omega}+C_0=0,
\end{align}
where
\begin{align}\label{c3}
	C_2=&\Delta_1+\Delta_2+\epsilon-i(\kappa_1+\kappa_2+\gamma_m),\notag\\
	C_1=&G_1^2 + G_2^2 + \kappa_1 (\kappa_2 + \gamma_m + i \Delta_2) +
	\kappa_2 [\gamma_m + i (\Delta_1 + \epsilon)]\notag\\
	-& [ \Delta_2 (\Delta_1 + \epsilon) -
	i\gamma_m (\Delta_1 + \Delta_2 + \epsilon)],\notag\\
	C_0=&i G_1^2 (\kappa_2 + i \Delta_2)+ i[G_2^2 +
	\gamma_m (\kappa_2 + i \Delta_2)]\notag\\
	\times& [\kappa_1 +
	i (\Delta_1 + \epsilon)].
\end{align}
Without the perturbation $\epsilon$, the considered system is a pseudo-Hermitian system when the condition in Eq.~(\ref{q11}) is satisfied. Using the pseudo-Hermitian condition, the coefficients of the characteristic equation (\ref{c3}) can be rewritten as
\begin{align}\label{c4}
	C_2=&(\eta-1)\Delta_1-\epsilon,\notag\\
	C_1=&(1+\eta+\eta^2)\kappa_2^2-G_1^2(1+\lambda^2)+\eta\Delta_1(\Delta_1+\epsilon)-i\eta\kappa_2\epsilon,\notag\\
	C_0=&G_1^2[(\Delta_1+\epsilon)\lambda^2-\eta\Delta_1]\notag\\
	-&\kappa_2(1+\eta)[(\Delta_1+\epsilon)\kappa_2-\eta^2\Delta_1\kappa_2-i\eta\Delta_1\epsilon].
\end{align}
Equation~(\ref{c2}) can be perturbatively expanded using a Newton-Puiseux series~\cite{Hodaei-2017}.
Considering only the first two terms, $\tilde{\Omega}\sim d_1 \epsilon^{1/3}+d_2\epsilon^{2/3}$, with the coefficients $d_1$ and $d_2$ being complex constants, results in
\begin{align}\label{c5}
	p_1\epsilon+p_{1/3}\epsilon^{1/3}+p_{2/3}\epsilon^{2/3}+p_{4/3}\epsilon^{4/3}+p_{5/3}\epsilon^{5/3}\notag\\
	+p_{6/3}\epsilon^{6/3}+p_{7/3}\epsilon^{7/3}+p_0=0,
\end{align}
where
\begin{align}\label{c6}
	p_0=&\kappa_2(1+\eta)(\Delta_1-i\eta\kappa_2)(\kappa_2-i\eta\Delta_1)\notag\\
	&+G_1^2[\Delta_1(\eta-\lambda^2)+i\kappa_2(1+\eta\lambda^2)],\notag\\
	p_1=&-d_1^3-2d_1d_2\Delta_1(\eta-1)\notag\\
	&+\kappa_2(1+\eta)(\kappa_2-i\eta\Delta_1)-\lambda^2G_1^2,\notag\\
	p_{1/3}=&d_1[\eta\Delta_1^2-\kappa_2^2(1+\eta+\eta^2)+(1+\lambda^2)G_1^2],\notag\\
	p_{2/3}=&d_1^2\Delta_1(1-\eta)+	p_{1/3}d_2/d_1,\notag\\
	p_{4/3}=&-3d_1^2d_2+d_2^2\Delta_1(1-\eta)+d_1\eta(\Delta_1-i\kappa_2),\notag\\
	p_{5/3}=&d_1^2-3d_1d_2^2+d_2\eta(\Delta_1-i\kappa_2),\notag\\
	p_{6/3}=&2d_1d_2-d_2^3,\notag\\
	p_{7/3}=&d_2^2.
\end{align}
For a specific case, we take the symmetric situation (i.e., $\eta=1$) of the pseudo-Hermitian COM system without $\mathcal{PT}$ symmetry, for instance, to illustrate the behavior of the eigenvalues near EP3. At EP3, we have $\lambda=1$, $G_1=2/\sqrt{3}\kappa_2$, and $\Delta_1=-1/\sqrt{3}\kappa_2$. Thus, the parameters in Eq.~(\ref{c6}) become
\begin{align}
	p_0=&0,~p_1=\frac{2}{3}(1+i\sqrt{3})-d_1^3,~p_{1/3}=0,~p_{2/3}=0,\notag\\
	p_{4/3}=&-\frac{-d_1}{\sqrt{3}}(1+i\sqrt{3})-3d_1^2d_2,\notag\\
	p_{5/3}=&d_1^2-\frac{d_2}{\sqrt{3}}(1+i\sqrt{3})-3d_1d_2^2,\notag\\
	p_{6/3}=&2d_1d_2-d_2^3,~p_{7/3}=d_2^2.
\end{align}
This indicates that the second and third terms in Eq.~(\ref{c5}) vanish. Forcing the coefficients of the first and fourth terms in Eq.~(\ref{c5}) to be zero, we obtain three sets of values for the coefficients $d_1$ and $d_{2}$, corresponding to the three eigenvalues:
\begin{align}
	(d_1,d_{2})=\Bigg(d_1^{(1)},d_{2}^{(1)}\Bigg),\Bigg(d_1^{(2)},d_{2}^{(2)}\Bigg),\Bigg(d_1^{(3)},d_{2}^{(3)}\Bigg),
\end{align}
where $d_1^{(1)}=1.10e^{i\pi/9}$, $d_1^{(2)}=-(0.550-0.953i)e^{i\pi/9}$, $d_1^{(3)}=-(0.550+0.953i)e^{i\pi/9}$, $d_{2}^{(1)}=-(0.175+0.303i)e^{-i\pi/9}$, $d_{2}^{(2)}=-(0.175-0.303i)e^{-i\pi/9}$, and $d_{2}^{(3)}=0.350e^{-i\pi/9}$. The bifurcations in the eigenvalues now acquire the following form
\begin{align}\label{c9}
\tilde{\Omega}_0\sim& d_1^{(1)}\epsilon^{1/3}+d_{2}^{(1)}\epsilon^{4/3},~	
\tilde{\Omega}_+\sim d_1^{(2)}\epsilon^{1/3}+d_{2}^{(2)}\epsilon^{4/3},\notag\\
\tilde{\Omega}_-\sim& d_1^{(3)}\epsilon^{1/3}+d_{2}^{(3)}\epsilon^{4/3}.
\end{align}
This indicates that the changes in the eigenvalues $\tilde{\Omega}_0$ and $\tilde{\Omega}_\pm$ follow the cube root of $\epsilon$,  because $\epsilon$ is very small. Such behavior can be characterized experimentally in the spectral domain by monitoring the resonant frequency splitting of, for example, $\tilde{\Omega}_+$ and $\tilde{\Omega}_0$~\cite{Hodaei-2017}, which can be expressed as
\begin{align}\label{c10}
	{\rm Re}[\Delta\tilde{\Omega}_{\rm EP3}]\sim&1.225\epsilon^{1/3}.
\end{align}
For the case of $\eta=2$ in the pseudo-Hermitian COM system without $\mathcal{PT}$ symmetry, results similar to those in Eqs.~(\ref{c9}) and~(\ref{c10}) can be obtained by repeating the above processes when the system  is driven to EP3 in the absence of the perturbation. For the case of the pseudo-Hermitian COM system with $\mathcal{PT}$ symmetry, the frequency splitting is the cubic-root dependence of the perturbation $\epsilon$, which was investigated previously~\cite{Hodaei-2017}. They also show that the frequency splitting is the square-root dependence of the perturbation $\epsilon$ when the system is operated near EP2~\cite{Hodaei-2017}.

\end{document}